\documentclass[9pt,twocolumn,twoside]{pnas-new}

\templatetype{pnasresearcharticle} 

\usepackage{graphicx, xcolor, multirow}
\usepackage{siunitx, amssymb, amsmath}

\title{On De Gennes Narrowing of Fluids Confined\\ 
at the Molecular Scale in Nanoporous Materials}

\author[a]{Wanda Kellouai}
\author[a,b]{Jean-Louis Barrat}
\author[c]{Patrick Judeinstein}
\author[a]{Marie Plazanet}
\author[a,1]{Benoit Coasne}

\affil[a]{ Univ. Grenoble Alpes, CNRS, LIPhy, F-38000 Grenoble, France}
\affil[b]{ Institut Laue-Langevin, 71 Avenue des Martyrs, 38042 Grenoble, France}
\affil[c]{ Université Paris-Saclay, CEA, CNRS, LLB, 91191 Gif-sur-Yvette, France}
\leadauthor{Kellouai et al.}

\significancestatement{Fluids confined in nanoporous materials display exotic structural, thermodynamic, and dynamical properties that challenge existing theories. These complex phenomena, which arise from the severe molecular confinement in nanopores, can be harnessed to design applications in storage, separation, and catalysis that are central in the energy/environmental crisis. Yet, such technological developments are hampered by our limited understanding of the molecular mechanisms that govern essential properties such as permeability (i.e. the fluid flow response induced by a pressure gradient). Here, we show that the structure and dynamics of a nanoconfined fluid are intimately linked through the so-called De Gennes narrowing concept. This powerful framework offers a robust formalism to rationalize the rich behavior of nanoconfined fluids from a simple microscopic picture.}

\authorcontributions{B.C.\ conceived the research. W.K.\ carried out the molecular simulations. W.K.\, B.C.\ and J.L.B.\ analyzed the data and developed the model. All authors contributed to the analysis of the results. 
B.C.\ wrote the manuscript with inputs from all authors.}
\authordeclaration{The authors declare no competing interests.}
\correspondingauthor{\textsuperscript{1} To whom correspondence should be addressed. E-mail: benoit.coasne@univ-grenoble-alpes.fr}


\begin{abstract}
Beyond well-documented confinement and surface effects arising from the large internal surface and severely confining porosity of nanoporous hosts, the transport of nanoconfined fluids remains puzzling by many aspects. With striking examples such as memory, \textit{i.e.} non-viscous, effects, intermittent dynamics and surface barriers, the dynamics of fluids in nanoconfinement challenges classical formalisms (\textit{e.g.} random walk, viscous/advective transport) -- especially for molecular pore sizes. In this context, while molecular frameworks such as intermittent brownian motion, free volume theory and surface diffusion are available to describe the self-diffusion of a molecularly confined fluid, a microscopic theory for the collective diffusion (\textit{i.e.} permeability) -- which characterizes the flow induced by a thermodynamic gradient -- is lacking. Here, to fill this knowledge gap, we invoke the concept of `De Gennes narrowing' which relates the wavevector-dependent collective diffusivity $D_0(q)$ to the fluid structure factor $S(q)$. First, using molecular simulation for a simple yet representative fluid confined in a prototypical solid (zeolite), we unravel an essential coupling between the wavevector-dependent collective diffusivity and the structural ordering imposed on the fluid by the crystalline nanoporous host. Second, despite this complex interplay with marked Bragg peaks in the fluid structure, the fluid collective dynamics is shown to be accurately described through De Gennes narrowing. Moreover, in contrast to the bulk fluid, departure from De Gennes narrowing for the confined fluid in the macroscopic limit remains small as the fluid/solid interactions in severe confinement screen collective effects and, hence, weaken the wavevector dependence of collective transport.
\end{abstract}

\dates{This manuscript was compiled on \today}
\doi{\url{www.pnas.org/cgi/doi/10.1073/pnas.XXXXXXXXXX}}

\begin{document}

\maketitle
\thispagestyle{firststyle}
\ifthenelse{\boolean{shortarticle}}{\ifthenelse{\boolean{singlecolumn}}{\abscontentformatted}{\abscontent}}{}

\dropcap{F}undamental research on fluids confined in nanoporous materials (pore size $\sim$ 1 -- 100 nm) keeps unveiling novel microscopic phenomena that are responsible for their unconventional structure, phase behavior, transport, and reactivity \cite{kavokine_2021,coasne_2013,huber_2015}. In particular, when the size of the confining porosity reaches the fluid molecular size ($\sigma$ $\leq$ nm), the specificities of the fluid/solid interaction combine with the ultra-confinement inside the material’s porosity to lead to a complex behavior involving both the physical and chemical features of the solid/fluid couple \cite{bocquet_2020, faucher_2019, sun_2020}. While the striking behavior of extremely confined fluids can be harnessed in nanotechnologies (\textit{e.g.} catalysis, separation, energy storage/conversion, depollution \cite{vandervoort, sholl_2016}), it often challenges existing theoretical formalisms \cite{kavokine_2021, bocquet_2010}. From a thermodynamic viewpoint, in addition to confinement-driven displacements in transition temperature/pressure and the appearance of surface-driven phase transitions, many aspects remain to be explored such as critical point shifts \cite{coasne_2013, evans_1986} or oversolubility in confined/vicinal fluids \cite{bratko_2008,ho_2013}. From a dynamical viewpoint, a broad collection of phenomena with complex underlying mechanisms can be observed including stop-and-go diffusion \cite{bousige_2021}, slip length and associated flow enhancements \cite{cottin-bizonne_2003,siria_2014}, but also non-memory effects leading to a breakdown of the viscous flow hypothesis \cite{falk_2015}. In fact, this rich transport behavior involves different microscopic pictures as fluid dynamics under given conditions in a nanoporous solid results from both individual (self-diffusion of a tagged molecule in its surrounding environment) and collective (involving cross-correlations from molecules with their neighbors) phenomena. As explained below, while the self-diffusivity of a nanoconfined molecule is relatively well-understood even in complex environments, the impact of severe confinement and large solid/fluid interactions at the host surface on collective dynamics remains to be fully explored \cite{bocquet_2020}.

Formally, despite the possible complexity of the free energy landscape in a given nanoporous material, the self-diffusivity of a nanoconfined fluid can be described using conventional formalisms such as Brownian motion (random walk) and the transition state theory \cite{bouchaud_1990,smit_2008}. As a result, even in nanoporous solids with complex pore morphologies and topologies, there exist simple statistical mechanics approaches which allow capturing diffusion processes in this class of materials including intermittent dynamical aspects (succession of residence and relocation times within the porosity) \cite{levitz_2005,levitz_2006}. Moreover, the impact of fluid density on the self-dynamics of the nanoconfined fluid can be rationalized using approaches such as free volume theories \cite{falk_2015} and surface diffusion models \cite{reed_1981}. Yet, even if the self-diffusion is an important descriptor of fluid transport in a nanoporous solid, it provides a restricted vision of the underlying dynamics by neglecting collective phenomena corresponding to fluid/fluid cross-contributions. In particular, while such cooperative effects in ultraconfined fluids can be small or negligible in some cases, the large differences often measured between the self and collective diffusivities provide evidence for their key role in transport in nanoporous materials \cite{smit_2008,bukowski_2021}. Yet, in contrast to self-diffusion, collective mechanisms involved in the overall transport of a nanoconfined fluid remain often treated at an empirical/observational level despite their central role in the permeability of the nanoconfined fluid and, more generally, in its response to any thermodynamic perturbation. 
In more detail, in the context of Onsager’s phenomenological theory of transport \cite{barrathansen}, the following  transport coefficients characterize the mass flow $\mathbf{J} = \rho \mathbf{v}$ -- defined as the fluid density $\rho$ multiplied by the flow velocity $\mathbf{v}$ -- as induced by a pressure gradient $\nabla P$, by a density/concentration gradient $\nabla \rho$ or by a chemical potential gradient $\nabla \mu$ \cite{smit_2008}: 
%
\begin{equation}\label{eq2}
\mathbf{J} = - \rho K \nabla P
\textrm{,} \hspace {3mm} 
\mathbf{J} = - D_\textrm{T} \nabla \rho
\textrm{,} \hspace {3mm} \textrm{and} \hspace {3mm} 
\mathbf{J} = - \rho \frac{D_0}{k_\textrm{B}T} \nabla \mu
\end{equation}
with $K$ the so-called permeance, $D_\textrm{T}$ the transport diffusivity, and $D_0$ the collective diffusivity. While these Onsager coefficients correspond to different physical situations as encountered in chemical engineering applications, they are related to each other by thermodynamic quantities amenable to simple experiments. For instance, $D_\textrm{T} \sim D_0 (\partial \mu / \partial \rho)_T$ where the proportionality coefficient can be inferred from the adsorption isotherm $\rho(\mu,T)$ at constant temperature $T$ as measured under static conditions \cite{smit_2008}. Similarly, the collective diffusivity $D_0$ and permeance $K$ are related to each other without any assumption through the Gibbs-Duhem equation at constant temperature $\textrm{d}P = \rho \textrm{d}\mu$, \textit{i.e.} $K = D_0/\rho k_\textrm{B} T$ \cite{falk_2015}. 

On the one hand, the thermodynamic consistency between the different transport coefficients allows probing different experimental flow situations that share the same fundamental molecular mechanisms. On the other hand, despite this simple, unifying framework, a microscopic theory to rationalize and predict the collective transport of fluids confined in nanoporous materials is still missing. Indeed, the molecular description of the impact of nanoconfinement and surface forces on collective diffusion $D_0$, transport diffusivity $D_\textrm{T}$, and permeance $K$ in ultraconfining pores remains to be established. Here, with the aim to fill this knowledge gap, we invoke the concept of `De Gennes narrowing' \cite{degennes_1959,wu_2018} to predict collective transport in a nanoconfined fluid from its microscopic structure within the confining porosity. While this simple theory is often put forward to explain qualitatively a slowdown (narrowing) observed in the dynamics at a given wavevector $q$ (or equivalently on a lengthscale $l \sim 1/q$) \cite{ruta_2014,banchio_2006,saito_2016}, it has never been considered and, \textit{a fortiori}, quantitatively assessed for a molecularly confined fluid (the closest example found in the literature corresponds to colloids in a narrow slit pore but the confinement is not at the molecular scale  \cite{nygard_2018}). Using a prototypical system consisting of a simple Lennard-Jones fluid confined in a well-defined, crystalline nanoporous solid (all-silica zeolite), we show here that De Gennes narrowing accurately describes the complex underlying dynamics leading to collective diffusivity and, hence, permeability in a nanoporous material with pores at the molecular scale. In particular, despite the prominent structural ordering imposed by the confining matrix on the fluid and the intermittent nature of the microscopic dynamics corresponding to subsequent adsorption/relocation steps, we find that the wavevector-dependent collective diffusivity $D_0(q)$ of the nanoconfined fluid can be rationalized from its structure factor $S(q)$.\\ 

\noindent \textbf{De Gennes narrowing and structure/dynamics coupling.} To introduce the concept of De Gennes narrowing, let us consider the density distribution at a time $t$ of a set of $N$ particles, $\rho(\textbf{r},t) = \sum_{i = 1,N} \delta(\textbf{r}-\textbf{r}_i(t))$. The dynamics of this system is characterized by the intermediate coherent scattering functions which describe the time relaxation $\tau_0(\textbf{q})$ of density fluctuations involving wavevectors $\textbf{q}$ \cite{barrathansen}: 
\begin{equation}
F_\textrm{coh}(\textbf{q},t) = \frac{\langle \rho(\textbf{q},t) \rho^\ast(\textbf{q},0) \rangle}{N S(\textbf{q})} 
\label{Free-energy} \end{equation}
where $\rho(\textbf{q},t) = \int \rho(\textbf{r},t) \exp[-i \textbf{q} \cdot \textbf{r}] \textrm{d}^3\textbf{r} = \sum_{i = 1,N} \exp[-i \textbf{q} \cdot \textbf{r}_i(t)]$ are the Fourier components of the density distribution while the $\ast$ indicates their conjugated quantity.  $S(\textbf{q}) =  \langle \rho_\textbf{q} \rho^\ast_\textbf{q} \rangle / N$ is the static structure factor and $\langle \cdot \cdot \cdot \rangle$ denotes ensemble average over many configurations of the system (this definition ensures that the intermediate coherent scattering functions are normalized, \textit{i.e.} $F_\textrm{coh}(\textbf{q},0) = 1$). 
Such coherent functions, which can be assessed using neutron scattering techniques for instance, allow estimating the relaxation time as $\tau_0(\textbf{q}) = \int_0^\infty F_\textrm{coh}(\textbf{q},t) \textrm{d}t$. 
To provide a simple analytical expression for $\tau_0(\textbf{q})$, let us consider the collective diffusivity $D_0$ and transport diffusivity $D_\textrm{T}$ as defined in Eqs. (1). By Combining these linear relations with the condition of local mass conservation, \textit{i.e.} $\partial \rho(\textbf{r},t)/\partial t + \nabla \cdot \textbf{J}(\textbf{r},t) = 0$, one obtains: 
%

\begin{equation}\label{eq3}
\frac{\partial \rho(\textbf{r},t)}{\partial t} - D_\textrm{T} \nabla^2 \rho(\textbf{r},t) = 0
\hspace {3mm} \textrm{and} \hspace {3mm}
\frac{\partial \rho(\textbf{r},t)}{\partial t} - \frac{\overline{\rho}D_0}{k_\textrm{B}T} \nabla^2 \mu(\textbf{r},t) = 0
\end{equation}
where $\overline{\rho} = N/V$ is the average number of molecules in the system having a volume $V$. 
By solving the first relation in Eq. (\ref{eq3}) in Fourier space (\textit{i.e.} $\partial \rho(\textbf{q},t)/\partial t + q^2 D_\textrm{T} \rho(\textbf{q},t) = 0$ with $q = |\textbf{q}|)$, one arrives at the general solution $\rho(\textbf{q},t) = \rho(\textbf{q},0) \exp[- D_\textrm{T} q^2 t]$ which leads to:  
\begin{equation}
F_\textrm{coh}(\textbf{q},t) = \exp[- t/\tau_0(\textbf{q})]  
\hspace {2mm} \textrm{with} \hspace {2mm} \tau_0(\textbf{q}) = \frac{1}{D_\textrm{T} q^2}
\label{eq4} \end{equation}

To introduce De Gennes narrowing, we now solve in Fourier space the second relation in Eq. (\ref{eq3}) by invoking the Fourier components $\mu(\textbf{q},t) = \int \mu(\textbf{r},t) \exp[-i \textbf{q} \cdot \textbf{r}] \textrm{d}^3\textbf{r}$  of the chemical potential, \textit{i.e.} $\partial \rho(\textbf{q},t)/\partial t +  q^2 \overline{\rho}D_0  \mu(\textbf{q},t)/k_\textrm{B}T = 0$. At this stage, we also express the free energy $\mathcal{F}[\overline{\rho}]$ of the system as a functional of the average density $\overline{\rho}$ which is the sum of all $\textbf{q}$ modes \cite{barrathansen,chaikinlubensky}: 
\begin{equation}
\mathcal{F}[\overline{\rho}] = \frac{1}{2} \frac{k_\textrm{B}T}{\overline{\rho}V} \sum_{\textbf{q}} \frac{\rho(\textbf{q}) \rho^\ast(\textbf{q})}{S(\textbf{q})} 
\label{eq5} \end{equation}
 Using this expression, we can use the functional derivative to express the Fourier components of the chemical potential as $\mu(\textbf{q},t) = \delta \mathcal{F}[\overline{\rho}] / \delta  \rho(\textbf{q},t) =  \rho(\textbf{q},t)/\overline{\rho}V k_\textrm{B}T S(\textbf{q})$. Upon inserting this expression in the Fourier transform of the second equation in Eqs. (\ref{eq3}), one obtains $\rho(\textbf{q},t) = \rho(\textbf{q},0) \exp[-D_0 q^2 t/S(\textbf{q})]$ which leads to:  
\begin{equation}
F_\textrm{coh}(\textbf{q},t) = \exp[- t/\tau_0(\textbf{q})]  
\hspace {2mm} \textrm{with} \hspace {2mm} \tau_0(\textbf{q}) = \frac{S(\textbf{q})}{D_0 q^2}
\label{eq6} \end{equation}
This is the essence of De Gennes narrowing which predicts that the collective dynamics at a given wavevector $\textbf{q}$ relaxes with a time constant $\tau_0(\textbf{q}) = S(\textbf{q})/D_0 q^2$. The physical picture behind this simple yet powerful concept is as follows. On the one hand, density fluctuations with a wavevector $\textbf{q}$ such that the structure factor $S(\textbf{q})$ is small relax fast [small $\tau_0(\textbf{q}$)] as they correspond to a large free energy. In other words, such characteristic fluctuations bring the system far from its thermodynamic equilibrium so that its response is fast (the thermodynamic equilibrium corresponds to a free energy minimum, i.e. a maximum in the structure factor according to Eq. (5)]. On the other hand, density fluctuations with a wavevector $\textbf{q}$ corresponding to a large structure factor $S(\textbf{q})$ relax slowly [large $\tau_0(\textbf{q})$] towards equilibrium as they maintain the system close to a free energy minimum. Taking the macroscopic limit (\textit{i.e.} $|\textbf{q}| \to 0$) of Eqs. (\ref{eq4}) and (\ref{eq6}) leads to  $D_\textrm{T} = D_0/S(0)$. As illustrated below, this expression is consistent with the well-known relationship $D_\textrm{T} = \Gamma D_0$ between $D_\textrm{T}$ and $D_0$ through the Darken or thermodynamic factor $\Gamma = \rho / k_\textrm{B}T \times (\partial \mu / \partial \rho)_T$. Besides this simple macroscopic limit, the general situation $|\textbf{q}| \neq 0$ corresponding to finite wavelengths $l = 2\pi/|\textbf{q}|$ is expected to lead to the following De Gennes narrowing effects. For all $\textbf{q}$, defining the $\textbf{q}$-dependent transport diffusivity as $D_\textrm{T}(\textbf{q}) = 1/\tau_0(\textbf{q}) q^2$, we predict from Eq. (6) the following scaling:
\begin{equation}\label{eq7}
D_\textrm{T}(\textbf{q}) = \frac{1}{\tau_0(\textbf{q}) q^2} \sim \frac{1}{S(\textbf{q})}
\end{equation}
In what follows, we test the validity of De Gennes narrowing for a molecularly confined fluid in a crystalline nanoporous material. First, we unravel the thermodynamics and structure of the nanoconfined fluid by discussing the adsorption isotherm at room temperature $n_\textrm{a}(P)$ in the light of the structure factor $S(\textbf{q})$. Second,  by  investigating the self-diffusivity of the nanoconfined fluid, we unveil  the intermittent nature of the  dynamics in nanoconfinement as the fluid alternates between adsorption (\textit{i.e.} residence) and diffusion (\textit{i.e.} relocation) steps. Third, we examine the $\textbf{q}$-dependent  collective dynamics to discuss the validity of De Gennes narrowing in extreme confinement.\\
  
\section*{Results}

\noindent \textbf{Fluid adsorption and structure in nanoconfinement.} As a simple yet representative example of extremely confined fluids, we consider methane adsorption and transport in a nanoporous crystalline material as illustrated in Fig. 1(a). As described in the Methods section, methane is described as a single Lennard-Jones sphere while the zeolite solid is chosen as all-silica silicalite-1 zeolite \cite{iza}. The nanoporous structure in this zeolite is made up of straight channels in the $y$ direction ($\sim$ 0.55 nm) that intersect with sinusoidal, i.e. zigzag, channels in the $xz$ plane ($\sim$ 0.51–0.55 nm). Using Grand Canonical Monte Carlo simulations \cite{frenkelsmit}, the number of adsorbed molecules $n_\textrm{a}(P)$ was first determined as a function of gas pressure $P$ at room temperature by setting the zeolite framework in contact with a fictitious fluid reservoir. As shown in Fig. 1(b), owing to the nanometer size of the cavities in the zeolite, the corresponding adsorption isotherm follows a Langmuir-like adsorption type (\textit{i.e.} without capillary condensation); $n_\textrm{a}$ increases very fast with pressure $P$ in the low pressure regime and then plateaus as the zeolite channels get filled with methane \cite{coasne_2013, deroche_2019}. Throughout this work, the following fluid loadings -- identified as color circles in Fig. 1(b) -- were selected as they cover the entire pressure/loading range: $n_\textrm{a} = $  6, 12, 16, and 20 CH$_4$ molecules per zeolite unit cell (uc).    

To unravel the structure of methane confined in the zeolite, Fig. 1(c) shows the structure factor of the confined fluid $S(\textbf{q}) = 1/N \langle \rho(\textbf{q}) \rho^\star(\textbf{q}) \rangle $ where we recall that $\rho(\textbf{q}) = \sum_{i =1,N} \exp[-i \textbf{q} \cdot \textbf{r}_i]$ (we underline that these data do not include the zeolite contribution as the sum is restricted to fluid particles). In more detail, Fig. 1(c) shows the structure factor for the different methane loadings $n_\textrm{a}$ considered here and for wavevectors aligned along the $x$ axis, \textit{i.e.} $q = \textbf{q} \cdot \textbf{e}_x$ ($\textbf{e}_x$ is the unit vector taken along the crystallographic axis $a$). For the sake of clarity, \textcolor{blue}{\textit{SI Appendix}, Fig. S1} shows the same data but in a linear scale with a zoom in the region around the main fluid correlation peak located at $q = q_\textrm{L}$. For each $n_\textrm{a}$, $S(q)$ reveals a dual nature with both liquid-like and solid-like features. On the one hand, nanoconfined methane displays a liquid-like structure with a marked but broad correlation peak centered around a wavevector $q_\textrm{L} \sim 1.8$ \AA$^{-1}$ corresponding to a typical distance $l = 2\pi/q_\textrm{L} \sim 3.5$ $\textrm{\AA}$ (a value of the order of the kinetic diameter of the fluid as expected). Upon decreasing the loading $n_\textrm{a}$, $q_\textrm{L}$ tends towards a lower value as the average distance between fluid molecules increases. Comparison in Fig. 1(c) with the structure factor obtained for the bulk fluid at a similar density than the largest loading considered here shows that this correlation peak is located at a similar wavevector $q$. This result indicates that the dense fluid structure within the zeolite possesses a similar nearest neighbor distance than its bulk counterpart. However, as shown in the zoom provided in \textcolor{blue}{\textit{SI Appendix}, Fig. S1}, this correlation peak is significantly broader for the bulk phase than for the confined phase. The narrower width $\Delta q$ for confined methane indicates that the coherence domain $L \sim 1/\Delta q$ in confinement is larger as the ordered nature of the confining host imposes long-range ordering. On the other hand, more strikingly, the structure factor for nanoconfined methane also displays crystalline features as the drastic impact of the zeolite crystalline structure also leads  to strong Bragg peaks (this interpretation was checked by verifying that the positions $q_\textrm{B}$ of these Bragg peaks are such that $q_\textrm{B} = 2\pi/\delta$ where $\delta$ is the distance along the $x$ direction between periodic channels in silicalite-1).\\

\begin{figure*}[htp]
  \centering
  \includegraphics[width=13.5cm]{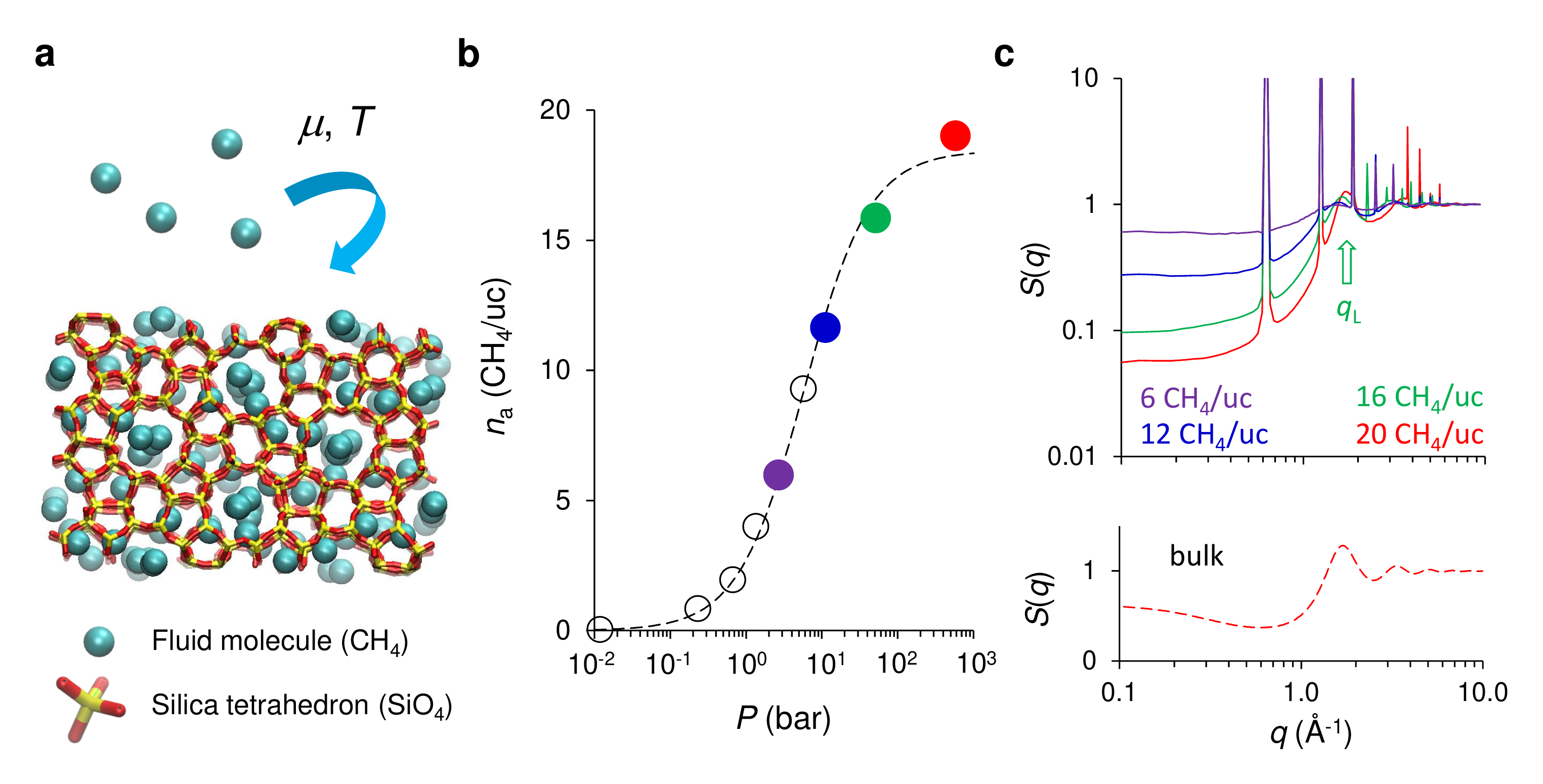}
  \caption{\textbf{Fluid adsorption and structure in a nanoporous material.} \textbf{a}, Typical molecular configuration of methane adsorbed in silicalite-1 zeolite at $T$ = 300 K and a pressure $P$ = 1000 bar. 
The blue spheres correspond to methane molecules while the yellow/red sticks denote the chemical bonds between Si and O atoms in the zeolite. 
This configuration was obtained by setting the zeolite in contact with a fluid reservoir imposing a chemical potential $\mu$ and temperature $T$ (with $\mu$ chosen to correspond to $P$ = 1000 bar). \textbf{b}, Methane adsorption isotherm at $T$ = 300 K in silicalite-1 zeolite. The adsorbed amount $n_\textrm{a}$ in CH$_4$ molecules per uc is expressed as a function of fluid pressure $P$. The color points correspond to the different loadings that are investigated in detail throughout this study. The dashed line, which is provided as a guide to the eye, corresponds to a fit of the data by a Langmuir adsorption isotherm. \textbf{c}, Structure factor $S(q)$ for bulk methane (bottom) and methane confined in silicalite-1 zeolite (top). 
For confined methane, each color denotes the fluid structure factor $S(q)$ obtained at the corresponding methane adsorbed amount as indicated in \textbf{b}.}
  \label{fig1}
\end{figure*}

\noindent	\textbf{Intermittent diffusion/residence transport.} 
With the aim to relate the microscopic structure and dynamics of nanoconfined methane, the self-diffusivity of methane was first investigated at different loadings $n_\textrm{a}$. As an example, Fig. 2(a) shows the mean square displacement $\langle \Delta x^2 \rangle$ along the $x$ direction as a function of time $t$ for $n_\textrm{a} \sim 20$ CH$_4$/uc. \textcolor{blue}{\textit{SI Appendix}, Fig. S2} also shows mean square displacements with a very similar behavior obtained when another direction and/or other loadings are considered. The data in Fig. 2(a) display the expected asymptotic regimes in the short time scale $\langle \Delta x^2 \rangle \sim t^2$ (ballistic regime) and in the long time scale $\langle \Delta x^2 \rangle \sim t$ (Fickian, \textit{i.e.} diffusive, regime). Between these two regimes, a short plateau is observed in $\langle \Delta x^2 \rangle$ for $\tau_\textrm{a} \sim 10^{-3}$ ns as molecules remain trapped in adsorption sites over a non-negligible time. Such behavior is characteristic of intermittent diffusion in nanoporous materials or in the vicinity of surfaces where the dynamics of adsorbing molecules consists of subsequent adsorption (residence) and diffusion (relocation) steps \cite{bousige_2021,levitz_2013}. Fig. 2(a) also shows a typical trajectory to illustrate such intermittent transport by displaying the center of mass of a given methane molecule taken every $\Delta t = 100$ fs. The interpretation above is confirmed by the analysis of the average velocity autocorrelation function $\langle v_x(t) v_x(0) \rangle$ shown in Fig. 2(b) for the same loading $n_\textrm{a}$. Besides back scattering observed as the velocity sign flips periodically in time, the typical decorrelation time taken as the time at which $\langle v_x(t) v_x(0) \rangle$ relaxes to zero roughly corresponds to the residence time $\tau_\textrm{a}$ observed in the mean square displacements.

The intermittent nature of fluid diffusion in the zeolite nanoporosity can be unraveled by determining the intermediate incoherent scattering functions $F_\textrm{inc}(\textbf{q},t)$ \cite{hansenmcdonald,calandrini_2011}. These quantities, which probe the displacement of individual particles corresponding to a wavevector $\textbf{q}$ over a time $t$, can be assessed using molecular simulation trajectories as: 
\begin{eqnarray}
F_\textrm{inc}(\textbf{q},t) = \frac{1}{N} \sum_{i=1}^N \rho_i(\textbf{q},t) \rho_i^\ast(\textbf{q},0) 
\nonumber
\\
= \frac{1}{N} \sum_{i=1}^N \exp[-i \textbf{q} \cdot (\textbf{r}_i(t)- \textbf{r}_i(0))]
\end{eqnarray}
where the second equality is obtained by invoking the instantaneous number density distribution for molecule $i$ in real space $\rho_i(\textbf{r},t) = \delta(\textbf{r}_i-\textbf{r}_i(t))$ into the Fourier transform $\rho_i(\textbf{q},t) = \int \rho_i(\textbf{r},t) \exp[-i \textbf{q} \cdot \textbf{r}]$. 
Fig. 2(c) shows the intermediate incoherent scattering function $F_\textrm{inc}(\textbf{q},t)$ for methane in silicalite-1 zeolite for different wavevectors $\textbf{q}$ taken along the $x$ direction so that $q = \textbf{q} \cdot \textbf{e}_x$. These data correspond to a methane loading $n_\textrm{a} = 20$ CH$_4$/uc but similar results were  obtained for other loadings. For small $q$ (large distances), the self-dynamics shows the expected diffusive regime corresponding to Fick’s second law, \textit{i.e.} $F_\textrm{inc}(\textbf{q},t) \sim \exp[-t/\tau(q)]$. As shown below, in this wavevector range, $\tau(q) = 1/D_s q^2$ where $D_s$ corresponds to the self-diffusivity obtained from the molecular trajectories in the long time scale. In contrast, for large $q$ (small distances), the self-dynamics shows a complex behavior as witnessed in the intermediate incoherent scattering functions $F_\textrm{inc}(\textbf{q},t)$ displaying both diffusion and adsorption phenomena. In more detail, diffusive processes correspond on the one hand to the decay in time of $F_\textrm{inc}(\textbf{q},t)$ with a characteristic time $\tau(q)$ that increases with decreasing $q$ (relaxation of density fluctuations over large distances occur on a long time scale). On the other hand, adsorption processes lead to the peak or shoulder observed at a characteristic time $\tau_\textrm{a}$. In contrast to diffusion, such a residence mechanism corresponds to localized motions in an adsorption site so that $\tau_\textrm{a}$ does not depend on $q$ (in this respect, molecular displacements in adsorption sites are equivalent to rotations which are also $q$-independent). 

\begin{figure*}[htp]
  \centering
  \includegraphics[width=13.5cm]{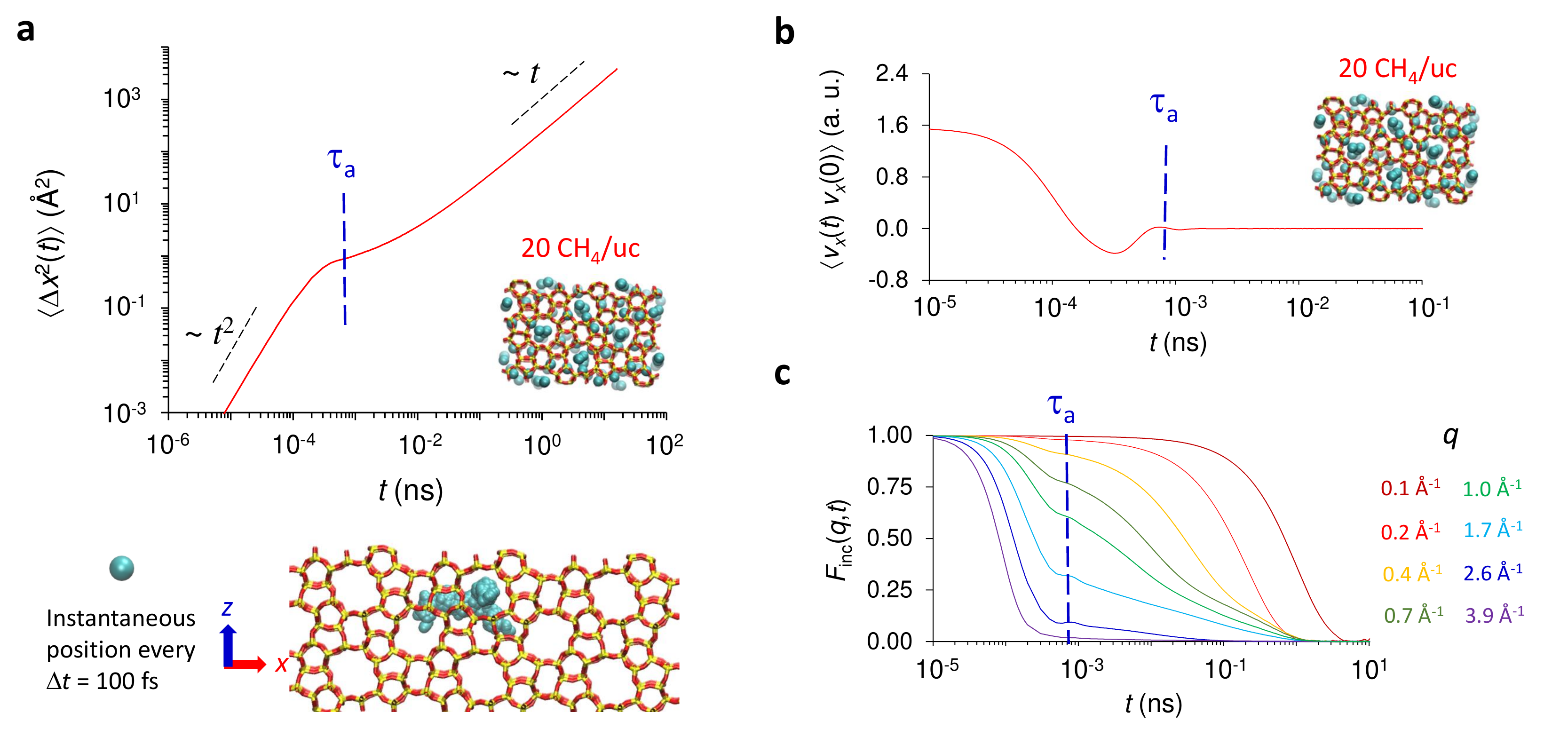}
  \caption{\textbf{Self-diffusion with underlying molecular intermittence.} \textbf{a}, Mean square displacements $\langle \Delta x^2 \rangle$ along the $x$ direction as a function of time $t$ for methane adsorbed in silicalite-1 zeolite at $T$ = 300 K. These data were obtained for a loading $n_\textrm{a} =$ 20 CH$_4$/uc. The dashed segments indicate the ballistic ($\langle \Delta x^2 \rangle \sim t^2$) and Fickian ($\langle \Delta x^2 \rangle \sim t$) regimes observed in the short and long time scales, respectively. The vertical blue dashed line denotes the characteristic time $\tau_a$ for a methane molecule to escape from an adsorption site (\textit{see text}). Diffusion in the severely confining zeolite porosity is illustrated using a typical molecular trajectory that displays the center of mass of a single methane molecule (cyan sphere) taken every $\Delta t = 100$ fs along the molecular dynamics trajectory. \textbf{b}, Velocity time autocorrelation function $\langle v_x(0) v_x(t) \rangle$ along the $x$ direction as a function of time $t$ for methane adsorbed in silicalite-1 zeolite at $T$ = 300 K. These data were obtained for the methane loading corresponding to $n_\textrm{a} \sim 20$ CH$_4$/uc. Like in \textbf{a}, the vertical red blue dashed line denotes the characteristic time $\tau_a$ for a methane molecule to escape from an adsorption site. \textbf{c}, Intermediate incoherent scattering function $F_\textrm{inc}(q,t)$ as a function of time $t$ for methane diffusion along the $x$ direction in silicalite-1 zeolite.  Like in \textbf{b}, these data were obtained for $n_\textrm{a} \sim 20$ CH$_4$/uc. As indicated in the graph, each curve corresponds to a $q$ vector taken parallel to the $x$ axis with a specific norm.}
  \label{fig2}
\end{figure*}

Having identified the intermittent nature of diffusion in nanoconfined methane, a more quantitative analysis of the data shown in Fig. 2 can be performed by considering the short and long time scales. As shown in \textcolor{blue}{\textit{SI Appendix}, Fig. S3}, the intermediate incoherent scattering functions at  small $q$ vectors can be reasonably fitted by a simple exponential decaying function \textit{i.e.} $F_\textrm{inc}(\textbf{q},t) = \exp[-t/\tau(\textbf{q})]$ (this result is confirmed by the fact that the same time constant $\tau(\textbf{q})$ is obtained when integrating $F_\textrm{inc}(\textbf{q},t)$ as shown in \textcolor{blue}{\textit{SI Appendix}, Fig. S4}).  As shown in Fig. 3(a), for the different loadings $n_\textrm{a}$,  $\tau(q)$ shows the expected scaling $\tau(q) = 1/D_s^x q^2$ where $D_s^x$ was taken from the slope of the mean square displacements in the long time range $D_s = 1/2 \lim_{t \to \infty} \textrm{d} \langle \Delta x^2(t) \rangle /\textrm{d}t$ (\textcolor{blue}{\textit{SI Appendix}, Fig. S5} illustrates that a similar self-diffusivity $D_s$ is obtained when considering the Green-Kubo formalism by integrating the velocity autocorrelation function $D_s^x  = \int_0^\infty \langle v_x(0) v_x(t) \rangle \textrm{d}t$). As expected, the Fickian regime holds for wavevectors $q << 2\pi/\lambda$ where $\lambda$ is the minimum distance that must be traveled by a molecule to reach Fickian diffusion in the mean square displacements $\langle \Delta x^2(t) \rangle  \sim t$. Overall, the data in Fig. 3(a) for the different loadings indicate that $D_s(n_\textrm{a})$ decreases with increasing $n_\textrm{a}$. \textcolor{blue}{\textit{SI Appendix}, Fig. S6} shows the self-diffusivity along the directions $x$ and $y$ as a function of $n_\textrm{a}$ which confirms that both $D_s^x$ and $D_s^y$ decrease as the loading $n_\textrm{a}$ increases. This behavior is characteristic of fluid diffusion in nanoporous materials such as zeolite as increased steric effects upon increasing $n_\textrm{a}$ hinder diffusion.

Besides Fickian diffusion in the long time regime, interpretation of the adsorption steps is less straightforward. In an attempt to rationalize the residence processes identified in the intermediate incoherent scattering functions, we propose the simple following model. We write the typical time $\tau_\textrm{a}$ needed to escape from an adsorption site as an activated process, \textit{i.e.} $\tau_\textrm{a}^\textrm{mod} = 1/\nu \exp[\Delta F/k_\textrm{B}T]$ where $\nu$ is the vibration frequency of the adsorbed molecule and $\Delta F$ is the activation energy that must be overcome to escape from the adsorption site. On the one hand, $\nu$ was estimated as the time between a maximum (positive) and a subsequent minimum (negative) in  the velocity autocorrelation function as observed in the short time range. On the other hand, $\Delta F$ was assessed by fitting the self-diffusivity $D_s$ as a function of temperature $T$ against an Arrhenius law, $D_s = D_s^0 \exp[-\Delta F/k_\textrm{B}T]$. As shown in Fig. 3(b), $\Delta F$ is found to decrease with the adsorbed amount $n_\textrm{a}$ with a decay that can be roughly described as  $\Delta F = a + b \exp[-n_\textrm{a}/c]$. This behavior shows that the activation energy required to induce  diffusive motion of the adsorbed molecules decreases with increasing fluid/fluid interactions. In other words, upon increasing the number of neighbors around an adsorbed molecule, collisions and intermolecular interactions with other confined molecules promote desorption and, hence, diffusion. $\Delta F$ is found to be larger but of the order of the thermal energy at room temperature $\Delta F \gtrsim RT$ so that the use of an Arrhenius plot should be considered with caution. Indeed, for such small energy barriers, the energy crossing rate $1/\nu$ and hence the diffusion coefficient depend on the exact shape of the free energy barrier \cite{kramers_1940,hanggi_1990}. While considering the shape in detail would allow determining accurately the rate $1/\nu$ \cite{festa_1978}, we note that the Arrhenius law is used here as an effective approach to describe from molecular dynamics trajectories the loading dependence of $D_s$ and $\tau_\textrm{a}(n_\textrm{a})$ 
(in other words, using a different phenomenological model from Arrhenius law would lead to very similar results). 
Comparison between the adsorption time as observed in the velocity autocorrelation function in Fig. 2(b), $\tau_\textrm{a}$, and the predicted adsorption time using the model above, $\tau_\textrm{a}^\textrm{mod}$, is shown in Fig. 2(c). Regardless of the adsorbed amount $n_\textrm{a}$ considered, the simple activated law for the adsorption time accurately describes the simulated data. In particular, both diffusion along the $x$ axis (zigzag channels) and along the $y$ axis (straight channels) is quantitatively predicted using this simple approach. On the one hand, the vibration frequency $\nu$, which is of the order of $2.4$ -- $2.8 \times 10^{12}$ Hz (corresponding to times in the range 350 -- 400 fs), is nearly independent of $n_\textrm{a}$ as expected for a single molecule adsorbed in a given site. On the other hand, $\nu$ and  $\tau_\textrm{a}$ were found to be dependent on the diffusion direction; due to the stronger confinement in the zigzag channels than in the straight channels, $\tau_\textrm{a}$ is larger along the $x$ direction than along the $y$ direction.\\

\begin{figure*}[htp]
  \centering
  \includegraphics[width=13.5cm]{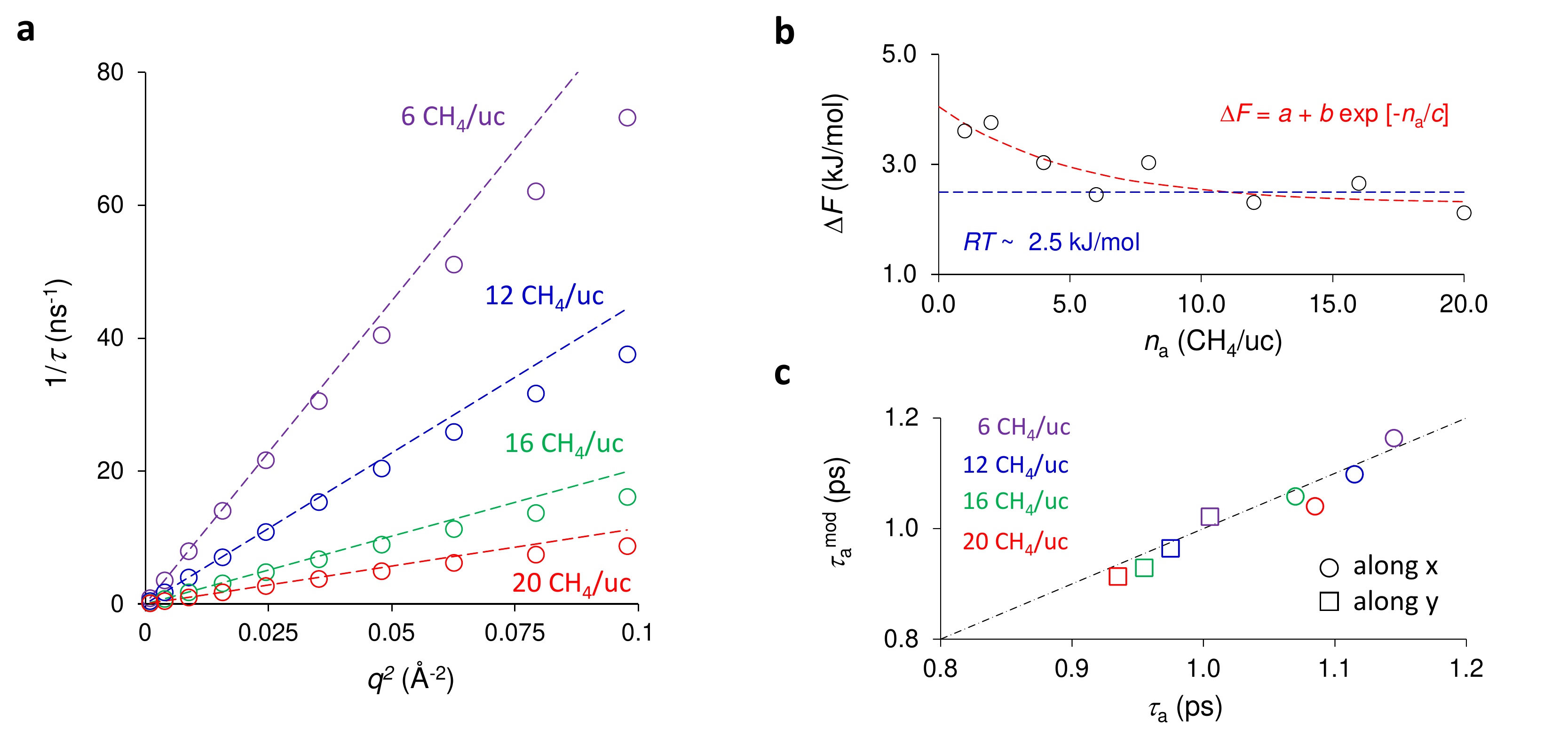}
  \caption{\textbf{Fickian regime with a characteristic residence time.} \textbf{a}, Characteristic time $\tau(q)$ as a function of wavevector $q$ for methane self-diffusion in silicalite-1 zeolite at $T$ = 300 K. Each color corresponds to a specific loading as indicated in the graph. The circles correspond to the characteristic times as obtained by fitting the intermediate incoherent scattering functions shown in Fig. 3(c) against an exponential function, \textit{i.e.} $F_\textrm{inc}(q,t) = \exp[-t/\tau(q)]$. The dashed lines correspond to the expected behavior in the Fickian regime when using the self-diffusivity $D_s$ assessed from the mean square displacements: $\tau(q) = 1/D_s q^2$. \textbf{b}, Free energy barrier $\Delta F$ as a function of loading $n_\textrm{a}$ for methane adsorbed in silicalite-1 zeolite at $T =$ 300 K. For each loading, these data were obtained by fitting the self-diffusivity measured at different temperatures against an Arrhenius law $D_s(T) \sim \exp[-\Delta F/k_\textrm{B}T]$. The blue line indicates the thermal energy at room temperature, $RT \sim 2.5$ kJ/mol while the red line is a fit of the data (circles) as a function of $n_\textrm{a}$ against $\Delta F = a + b \exp[-n_\textrm{a}/c]$ (\textit{see text}). \textbf{c}, Predicted escape time $\tau_\textrm{a}^\textrm{mod}$ from an adsorption site in silicalite-1 zeolite as a function of the observed time $\tau_\textrm{a}$. Each color corresponds to a specific methane loading as indicated in the graph. The circles and squares correspond to escape times along the $x$ and $y$ directions, respectively. While $\tau_\textrm{a}$ is determined from the residence time observed in the intermediate incoherent scattering function $F_\textrm{inc}(q,t)$, $\tau_\textrm{a}^\textrm{mod}$ is predicted using a simple activated model $\tau_\textrm{a}^\textrm{mod} \sim 1/\nu \exp[\Delta F/k\textrm{B}T]$ with $\Delta F$ taken from data in panel \textbf{b} (\textit{see text}).}
  \label{fig3}
\end{figure*}

\noindent	\textbf{Thermodynamics/structure consistency.} The collective diffusion and permeability of methane through the zeolite was first investigated by considering the macroscopic regime (small wavevector $\textbf{q}$). As shown in the inset of Fig. 4(a), the intermediate coherent scattering functions for $q = |\textbf{q}| < 0.4$ \AA$^{-1}$ shows the expected behavior with $F_\textrm{coh}(\textbf{q},t) = \exp[-t/\tau_0(\textbf{q})]$ for all methane loadings $n_\textrm{a}$. Moreover, upon fitting the data, we found that $\tau_0(\textbf{q}) = 1/D_\textrm{T}q^2$ with $D_\textrm{T}$ equal to the macroscopic transport diffusivity. With the aim to confirm this interpretation, recalling that the collective and transport diffusivities are linked to each other as $D_\textrm{T} = D_0/S(0)$ in the macroscopic limit $|\textbf{q}| \to 0$, we also determined $D_0$ as follows. Using non-equilibrium molecular dynamics as described in the Methods section, we performed simulations under different chemical potential gradients $\nabla \mu$. As shown in \textcolor{blue}{\textit{SI Appendix}, Fig. S7}, we found that the induced molecular flow $\textbf{J}$ is proportional to $-\nabla \mu$ provided the latter remains small enough (it was also checked that the induced flow rate $\langle v \rangle$ is very small compared to the thermal velocity $v_\textrm{th} \sim \sqrt{k_\textrm{B}T/m}$). The dashed line in Fig. 4(a) shows the transport diffusivity $D_\textrm{T} = D_0/S(0)$ predicted from the collective diffusivity $D_0$ inferred by fitting the non-equilibrium molecular dynamics data against a linear behavior, \textit{i.e.} $\langle v \rangle = - D_0/k_\textrm{B}T$ [as for $S(0)$, it is directly obtained from the data in Fig. 1(c)]. As can be seen in Fig. 4(a), the collective diffusivities $D_0$ determined at different $n_\textrm{a}$ are in good agreement with the transport diffusivities $D_T$ obtained from the intermediate coherent scattering functions. Such a comparison was extended to the $y$ direction for the different loadings in Fig. 4(b) where we plot $D_\textrm{T}$ as a function of $D_0/S(0)$.  We also performed this consistency check for the maximum loading $n_\textrm{a} \sim 20$ CH$_4$/uc in a rigid (\textit{i.e.} non-thermalized) zeolite. As can be seen in Fig. 4(b), regardless of the loading, transport direction, and thermalization of the nanoporous material, the transport $D_\textrm{T}$ and collective $D_0$ diffusivities are found to be consistent. \textcolor{blue}{\textit{SI Appendix}, Fig. S6} shows the collective diffusivity $D_0$ along the $x$ and $y$ directions as a function of $n_\textrm{a}$. Like for the self-diffusivity, both $D_s^x$ and $D_s^y$ decrease with increasing $n_\textrm{a}$ due to enhanced steric repulsion limiting transport. Moreover, the ratio $D_0/D_s$ increases as collective effects in the transport properties of the nanoconfined fluids become more prominent upon increasing the loading.

As already stated, considering that the transport diffusivity $D_\textrm{T}$ and collective diffusivity $D_0$ characterize the molecular flow induced by a density gradient and chemical potential gradient, it is straightforward to show that: 
\begin{equation}\label{eq9}
D_\textrm{T} = \rho D_0/k_\textrm{B}T  \bigg (\frac{\partial \mu}{\partial \rho}\bigg )_T =  D_0 \bigg(\frac{\partial \ln f}{\partial \ln \rho}\bigg)_T
\end{equation}
where the subscript $T$ indicates that the partial derivative is taken at constant temperature. The second equality in the above equation is obtained by invoking the fugacity defined as $\mu = k_\textrm{B}T \ln (f\Lambda^3/k_\textrm{B}T)$ ($\Lambda$ is the De Broglie thermal wavelength). We note that the fugacity as expressed here in Pa corresponds to the pressure of the $n_\textrm{a}$ molecules if they behaved as an ideal gas phase. While Eq. (\ref{eq9}) shows that $D_\textrm{T}$ and $D_0$ are related through a thermodynamic factor, i.e. $\Gamma = (\partial \ln f/\partial \ln \rho)_T$, the previously introduced relationship $D_\textrm{T} = D_0/S(0)$ indicates that these two transport coefficients are linked through the structural quantity $S(0)$ at vanishing wavevectors $q \to 0$. By noting that $S(0) = \rho k_\textrm{B}T \chi_T$ where $\chi_T = 1/\rho(\partial  \rho/ \partial P)_T$ is the isothermal compressibility, the use of Gibbs-Duhem equation at constant temperature $\textrm{d}P = \rho \textrm{d}\mu$ allows us to show $S(0) = k_\textrm{B}T/\rho   \times  (\partial \rho / \partial \mu)_T$ and, hence, to recover $D_\textrm{T} = D_0/S(0) = \rho D_0/k_\textrm{B}T \times (\partial \mu/\partial \rho)_T$. This thermodynamic/structure equivalence between $D_\textrm{T}$ and $D_0$ was verified by assessing the thermodynamic factor $\Gamma$ as a function of loading $n_\textrm{a}$ from the slope of the adsorption isotherm reported in Fig. 1. As illustrated in Fig. 4(c), $S(0)$ and $\Gamma$ were found to obey the correct behavior at all loadings with $S(0) = \Gamma^{-1}$. Beyond this expected relationship, it is interesting to note that the equivalence between $D_\textrm{T}$ and $D_0$ apply to all directions (\textit{i.e.} $x$, $y$, and $z$)  since both $S(0)$ and $\Gamma$ are scalar  quantities that do not depend on the specific transport direction being considered. On the one hand, such an isotropic behavior is trivial for $\Gamma$ since the adsorption isotherm at a given temperature $T$ is a scalar thermodynamic quantity that only depends on the material/fluid couple. On the other hand, for the structure factor, despite drastically different molecular distributions along the three directions of space, $S(0) = \lim_{|\textbf{q}|\to 0} S(|\textbf{q}|) = \rho   k_\textrm{T} \chi_T$ is direction independent as the confined phase behaves as a fluid so that $\rho$ and $\chi_T$ are isotropic, scalar quantities.\\

\begin{figure*}[htp]
  \centering
  \includegraphics[width=13.5cm]{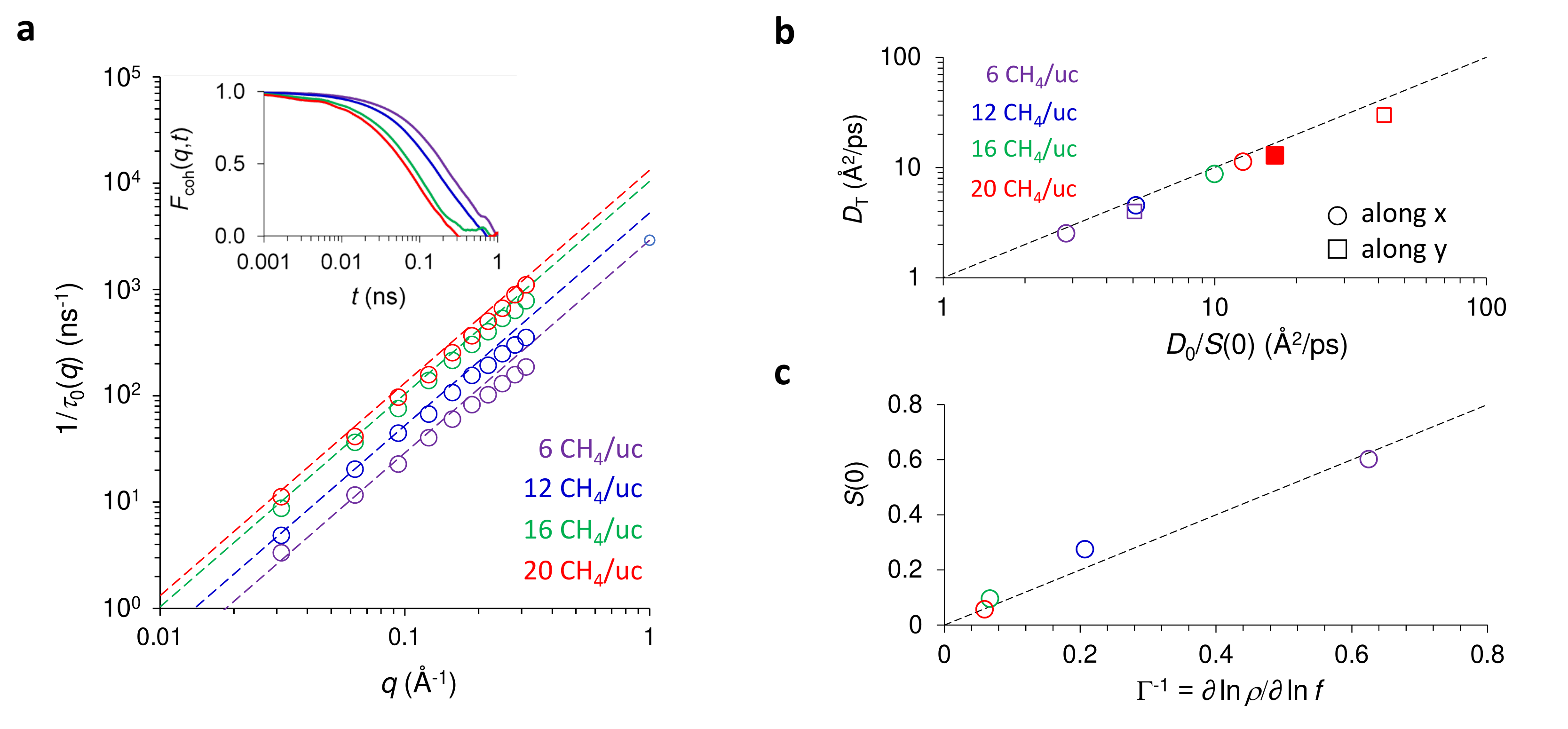}
  \caption{\textbf{Consistency between structure, thermodynamics and dynamics in collective transport.} \textbf{a}, Characteristic time $\tau_0(q)$ as a function of wavevector $q$ for methane collective diffusion in silicalite-1 zeolite at $T$ = 300 K. Each color corresponds to a specific loading $n_\textrm{a}$ as indicated in the graph. The circles correspond to the characteristic times obtained by fitting the intermediate coherent scattering functions $F_\textrm{coh}(q,t)$ shown in the insert against an exponential function, \textit{i.e.} $F_\textrm{coh}(q,t) = \exp[-t/\tau_0(q)]$. The dashed lines correspond to the expected behavior in the small $q$ wavevector regime when using the collective diffusivity $D_0$ assessed from non-equilibrium molecular dynamics. As explained in the text, the latter is obtained by probing the linear response between a chemical potential gradient $\nabla \mu$ and the induced molecular flow $J$, \textit{i.e.} $J = - \rho D_0/k_\textrm{B}T \nabla \mu$. \textbf{b}, Consistency check between the transport diffusivity $D_\textrm{T}$ and  the collective diffusivity $D_0$ which are formally related through the structure factor at vanishing wavevectors, \textit{i.e.} $S(0) = \lim_{q \to 0} S(q)$. Each color corresponds to a different loading $n_\textrm{a}$ as indicated in the graph. The circles and squares correspond to the transport coefficient $D_\textrm{T}$ along the $x$ and $y$ directions, respectively. The open symbols are for the flexible zeolite while the close symbol is a test performed for a rigid zeolite. \textbf{c}, Verification of the formal equivalence between the structure factor at vanishing wavevectors $q$, $S(0) = \lim_{q \to 0} S(q)$ and the so-called thermodynamic or Darken factor $\Gamma^{-1} = (\partial \ln \rho / \partial \ln f)_T$ as measured from the adsorption isotherm reported in Fig. 1(a). Like for the other panels, each color corresponds to a different methane loading $n_\textrm{a}$ in silicalite-1 zeolite.}
  \label{fig4}
\end{figure*}

\noindent	\textbf{De Gennes narrowing in molecularly confined fluids.} With the aim to assess the ability of \textit{De Gennes narrowing} concept to describe the transport of fluids confined in a nanoporous material, we now analyze the microscopic behavior of the fluid collective dynamics for different wavevectors in the light of the structure factor $S(q)$. Fig. 5(a) shows  the time evolution of the intermediate coherent scattering function $F_\textrm{coh}(q,t)$ for the loading $n_\textrm{a} = 20$ CH$_4$/uc taken at different wavevectors along the $x$ axis, \textit{i.e.} $q = \textbf{q} \cdot \textbf{e}_x$. We also show in Fig. 5(b) the structure factor $S(q)$ for the same loading as a function of $q = \textbf{q} \cdot \textbf{e}_x$. In addition to providing microscopic details about the structure of the nanoconfined fluid, $S(q) \sim \langle \rho(q,0) \rho^\ast(q,0) \rangle$ also corresponds to the normalization constant at $t = 0$ for the intermediate coherent scattering function $F_\textrm{coh}(q,t) \sim \langle \rho(q,t) \rho^\ast(q,0) \rangle$. As illustrated in Fig. 5(a) for the loading $n_\textrm{a} = 20$ CH$_4$/uc, $F_\textrm{coh}(q,t)$ shows the expected decaying behavior towards zero for wavevectors that differ from Bragg peaks, \textit{i.e.} $q \neq q_\textrm{B}$. As for $q$ chosen among the Bragg peaks, \textit{i.e.} $q =  q_\textrm{B}$, we observe that $F_\textrm{coh}(q,t)$ does not relax to zero but towards a finite, constant value. In fact, for all $q$,  the intermediate coherent scattering function converges to $\lim_{t \to \infty} \langle \rho(q,t) \rho^\ast(q,0) \rangle = |\langle \rho(q) \rangle|^2$ with $|\langle \rho(q) \rangle| = 0$ for $q \neq q_\textrm{B}$ and  $|\langle \rho(q) \rangle| \neq 0$ for $q = q_\textrm{B}$. This is due to the fact that the crystalline structure of the host zeolite prescribes a similar underlying structure which leads to the strong Bragg peaks  $S(q_\textrm{B})$. As a result, while $\langle \rho(q) \rangle \sim 0$  for $q \neq  q_\textrm{B}$, $\langle \rho(q) \rangle \neq 0$ for $q = q_\textrm{B}$. In order to correct for such a static structural effect corresponding to the fingerprint of the host zeolite on the nanoconfined fluid,  we redefine the intermediate coherent scattering functions as $\tilde{F}_\textrm{coh}(q,t) \sim  \langle \delta \rho(q,t) \delta \rho^\ast(q,0)  \rangle $ where $\delta \rho(q,t) = \rho(q,t) - \langle  \rho(q) \rangle$. Similarly, we can redefine the structure factor to retain only the the dynamical, \textit{i.e.} fluctuating, contribution as $\tilde{S}(q) = 1/N \langle \delta \rho(q) \delta \rho^\ast(q)   \rangle$. As shown in Fig. 5(a) for the intermediate coherent scattering functions and in Fig. 5(b) for the dynamic structure factor,  this correction only modifies the data for $q = q_\textrm{B}$ as $\langle \rho(q) = 0\rangle$ for  $q \neq q_\textrm{B}$. Interestingly, despite the very strong Bragg peaks observed in $S(q)$, it is observed that $\tilde{S}(q)$ for $q = q_\textrm{B}$ is a local minimum. This unexpected result indicates that, due to the very strong correlations imposed by the confining host on the fluid microscopic organization, fluid-fluid correlations are weak at these specific wavevectors [see zooms in the Bragg peak regions as provided in Fig. 5(b)]. In other words, the density correlations in the fluid molecular distribution within the porosity for $q = q_\textrm{B}$ are mostly governed by the periodic structure of the host zeolite with very limited impact of the fluid-fluid correlations. 

Like for the intermediate incoherent scattering functions $F_\textrm{inc}(q,t)$, the coherent part shows a rapid decay at a time $\tau_0(q)$ which increases upon decreasing $q$ since relaxation on large lengthscales occurs on longer times. Moreover, the collective behavior in the confined dynamics as scrutinized with the $\tilde{F}_\textrm{coh}(q,t)$ functions also displays the signature of adsorption/diffusion  intermittent dynamics -- \textit{i.e.}  a strong decay in time combined with localized motions at a time scale that is only weakly $q$-dependent. It was checked that the latter feature in the coherent dynamics, which corresponds to adsorption stages in well-localized sites, coincides with that observed in the $F_\textrm{inc}(q,t)$ functions. As a result, considering that these localized motions correspond to an individual, \textit{i.e.} self, property of the confined fluid dynamics, it was not considered in our assessment of the validity of De Gennes narrowing applied to a nanoconfined fluid. In practice, as described in detail below, we fitted the diffusive part to obtain the relaxation time $\tau_0(q)$ as a function of $q$. While this suggests that we discard the residence phenomenon from collective transport, we emphasize that the diffusive dynamics as revealed by the function $\tau_0(q)$ and, hence, $D_\textrm{T}(q)= 1/\tau_0(q)q^2$, does reflect adsorption since it is also affected by the intermittent nature of the confined dynamics. In particular, in the long time scale and/or small wavevectors, as already discussed when commenting Fig. 4, $F_\textrm{coh}(q,t)$ exhibits the typical time decay with a constant that allows reproducing the macroscopic transport coefficients as obtained using non-equilibrium molecular dynamics. Fig. 5(c) shows the reciprocal of the characteristic relaxation time $1/\tau_0(q)$ as a function of $q$ as estimated from the intermediate coherent scattering functions $\tilde{F}_\textrm{coh}(q,t)$. Different approaches were considered: (1) fit against an exponential decaying function $\tilde{F}_\textrm{coh}(q,t) \sim \exp[-t/\tau_0(q)]$, (2) fit against a stretched exponential decaying function $\tilde{F}_\textrm{coh}(q,t) \sim \exp[-(t/\tau(q))^\beta]$ with the characteristic time then estimated as $\tau_0(q) = \tau(q) \beta^{-1} \Gamma^\ast(\beta^{-1})$ [$\Gamma^\ast(x)$ is the gamma function] \cite{nygard_2018}, and (3) $\tau_0(q) = \tau_e$ where $\tau_e$ is the time at which $\tilde{F}_\textrm{coh}(q,t) = 1/e$. For $q \lesssim 0.25$ \AA$^{-1}$, a simple exponential function was found to fit very well the data. As a result, for this asymptotic regime, the inferred  $\tau_0(q)$ is very close to that obtained as $\tilde{F}_\textrm{coh}(q,\tau_0(q)) = 1/e$ (consistently, we also found that $\tau_0(q)$ is very close to that obtained through the integral $\int_0^\infty \tilde{F}_\textrm{coh}(q,t) \textrm{d}t$). As shown in Fig. 5(c), this expected diffusive regime is confirmed by the fact that $1/\tau_0(q) \sim D_\textrm{T}q^2$ in this low $q$ range. For $q > 0.25$ \AA$^{-1}$, the situation is more complicated as (1) the diffusive decay in $\tilde{F}_\textrm{coh}(q,\tau_0(q))$ is no longer a simple exponential function and (2) the overlap between localized and diffusive motions prevents the use of a simple function. As a result, we compared the use of a stretched exponential function and the simple estimate corresponding to $\tau_e$ to assess the typical relaxation time $\tau_0(q)$. While these two approaches provide comparable relaxation constants $\tau_0(q)$ in Fig. 5(c), the use of $\tau_e$ was found to provide an appropriate description of the diffusion contribution in the $\tilde{F}_\textrm{coh}(q,t)$ functions for $q > 0.25$ \AA$^{-1}$. In particular, despite being a simple and crude description of the decay in $\tilde{F}_\textrm{coh}(q,t)$, we found that $\tau_e$  provides a robust description of the time constant involved in the collective dynamics that is less prone to arbitrary choices (fitting range, statistical uncertainty, etc.).

\begin{figure*}[htp]
  \centering
  \includegraphics[width=13.5cm]{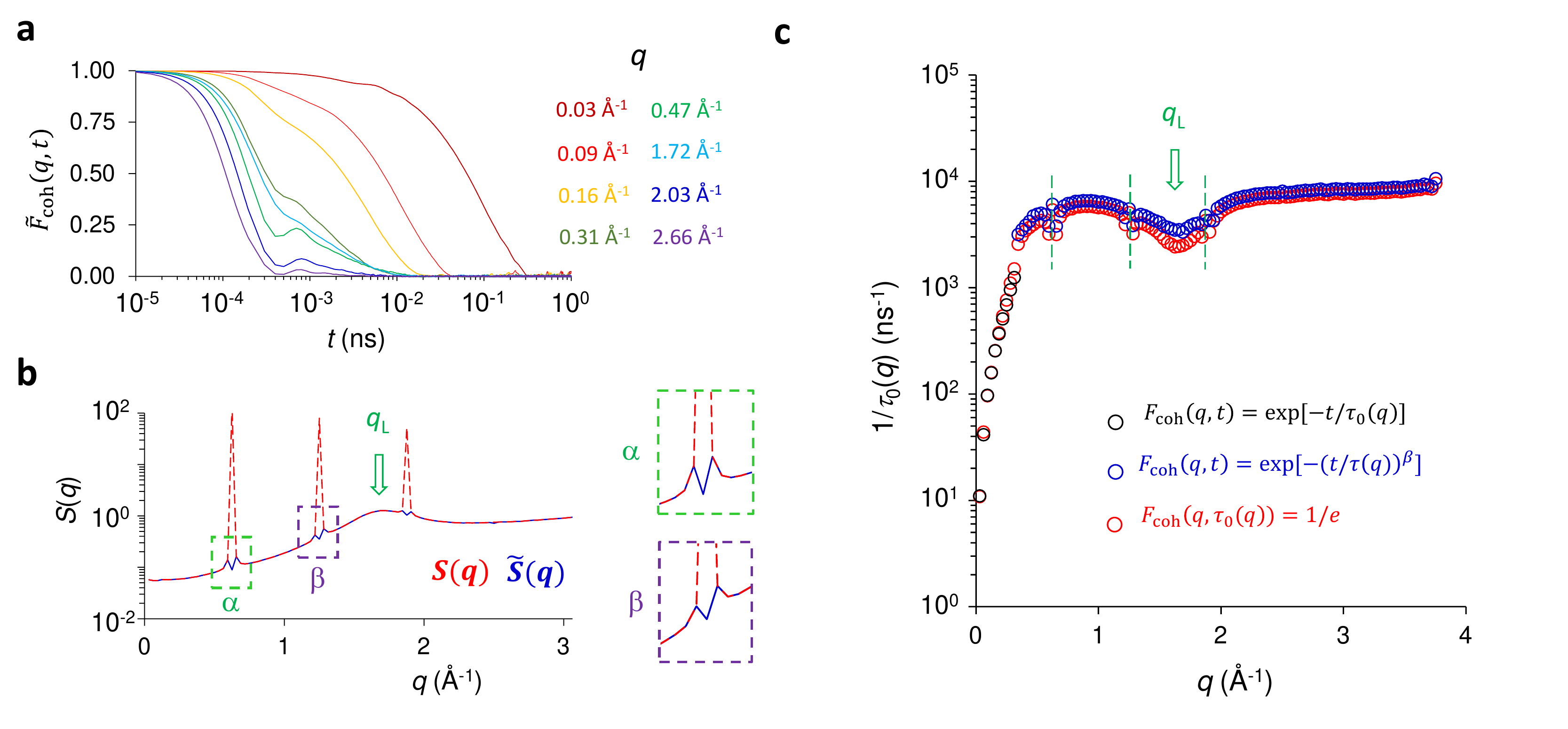}
  \caption{\textbf{Intermediate scattering functions and characteristic times for collective transport.} \textbf{a}, Intermediate coherent scattering functions $F_\textrm{coh}(q,t)$ as a function of time $t$ for methane adsorbed in silicalite-1 zeolite with a loading of $\sim 20$ CH$_4$/uc. Each color corresponds to a different wavevector $q$ as indicated in the graph. \textbf{b}, Structure factor $S(q)$ for methane adsorbed in silicalite-1 zeolite with a loading of $\sim 20$ CH$_4$/uc. The red dashed line indicates the nominal structure factor $S(q) = 1/N \langle \rho(q) \rho^\ast(q) \rangle$. The blue line, which only departs from $S(q)$ for wavevectors $q$ corresponding to Bragg peaks $q_\textrm{B}$, shows the fluctuating part $\tilde{S}(q) = 1/N \langle \delta \rho(q) \delta \rho^\ast(q) \rangle$ where  $\delta \rho(q) = \rho(q) - \langle\rho(q) \rangle$ (\textit{see text}). As can be seen in the zooms provided for regions $\alpha$ and $\beta$, despite the very strong peak at Bragg conditions, $\tilde{S}(q)$ displays a minimum for $q = q_\textrm{B}$. \textbf{c}, Characteristic time $\tau_0(q)$ as a function of wavevector $q$ for methane collective diffusion in silicalite-1 zeolite at $T$ = 300 K (loading  $n_\textrm{a} \sim 20$ CH$_4$/uc).  The black circles correspond to $\tau_0(q)$  as obtained by fitting the intermediate coherent scattering functions against a simple exponential function, \textit{i.e.} $F_\textrm{coh}(q,t) = \exp[-t/\tau_0(q)]$. The blue circles correspond $\tau_0(q)$  as obtained by fitting the intermediate coherent scattering functions against a stretched exponential function, \textit{i.e.} $F_\textrm{coh}(q,t) = \exp[-(t/\tau(q))^\beta]$ [in that case the corrected decaying time is given by $\tau_0(q) = \tau(q) \beta^{-1} \Gamma^\star(\beta^{-1})$ where $\Gamma^\star(x)$ is the gamma function. The red circles correspond to the characteristic time defined as $F_\textrm{coh}(q,\tau_0(q)) = 1/e$. The green down arrow indicates the position of the liquid correlation peak $q_\textrm{L}$ in the structure factor $S(q)$ while the dashed vertical segments correspond to the positions $q_\textrm{B}$ in the structure factor $S(q)$.}
  \label{fig5}
\end{figure*}

Fig. 6(a) shows $D_\textrm{T}(q) = 1/\tau_0(q) q^2$ as a function of $q$ for methane confined in the zeolite with a loading $n_\textrm{a} = 20$ CH$_4$/uc. Qualitatively, $D_\textrm{T}(q)$ displays De Gennes narrowing as it shows marked features that closely follow the variations observed in the structure factor $S(q)$ taken under the same conditions. As an indication of the collective nature of the variations observed in $D_\textrm{T}(q)$, we remark that the $q$-dependent self-diffusivity $D_s(q)$ in Fig. 6(a) does not display the same behavior. $D_\textrm{T}(q)$ shows narrowing (lowering) for $q$ around $q_\textrm{L}$ where significant fluid-fluid correlations are observed. Moreover, $D_\textrm{T}(q)$ also shows marked variations for wavevectors near the Bragg peaks $q \sim q_\textrm{B}$. 
To quantitatively assess the validity of De Gennes narrowing, \textit{i.e.} $D_\textrm{T}(q) = 1/\tau_0(q) q^2 \sim 1/\tilde{S}(q)$, we show  $D_\textrm{T}(q) \times \tilde{S}(q)$ as a function of $q$ in Fig. 6(c). We also report the data obtained for bulk methane taken at the same temperature $T$ and density $\rho$. \textcolor{blue}{\textit{SI Appendix}, \textit{SI Text}} presents in detail the analysis for the bulk fluid which leads to the corresponding data shown in Fig. 6(c). For both the bulk and confined fluids, the plateau observed for $q$ around the fluid-fluid correlation peak $q_\textrm{L}$ validates the concept of De Gennes narrowing, \textit{i.e.} $D_\textrm{T}(q) \sim 1/\tilde{S}(q)$ [see the shaded areas in Fig. 6(c)]. Interestingly, as shown by the big red crosses in Fig. 6(c), considering the static structure factor $S(q)$ instead of the fluctuating part $\tilde{S}(q)$ does not allow rationalizing the $q$-dependent collective dynamics. Indeed, the large Bragg peaks in $S(q)$ do not govern the confined fluid dynamics at the corresponding wavevectors but simply reveal the strong fingerprint of the host confining structure. De Gennes narrowing can also be pictured in the inset of Fig. 6(c) which shows $D_\textrm{T}(q)$ as a function of $S(q)$ in a log-log scale. 

At large $q$, both the confined and bulk fluid dynamics depart from De Gennes narrowing as $D_\textrm{T}(q)$ decreases with increasing $q$ while $S(q)$ remains nearly constant. Such departure in this high $q$ range is due to the fact that the Onsager expressions given in Eqs. (1) are only valid at sufficiently large time and length scales. Typically, while the size domain $l$ over which these expressions can be applied is system dependent, it is reasonable to assume that it must be at least equal or larger than the typical molecular size $\sigma$ as observed here: $l \sim 2\pi/q \sim 3$ \AA $ $ ($q \sim 1.8$ -- $2.2$ \AA$^{-1}$). Moreover, while De Gennes narrowing applies on a similar $q$-range for the bulk and confined fluids, it holds to even larger wavevectors for the confined fluid. This result is believed to be due to the fact that the confined fluid has a more pronounced structure which drives its dynamics to even smaller lengthscales (See \textcolor{blue}{\textit{SI Appendix}, Fig. S1} where we compare the structure factors $S(q)$ for the bulk and confined fluids under similar temperature and density conditions). As for the small $q$ range, while De Gennes narrowing applies to smaller wavevectors for the confined fluid, the data in Fig. 6(c) show departure for both the bulk and confined fluids for $q < 0.7$ -- $0.8$ \AA$^{-1}$. This lower boundary in the $q$-range corresponds to a mesoscopic, \textit{i.e.} supramolecular, length $l \sim 8$ -- 9 \AA$^{-1}$. As discussed in detail below, in this range, the assumption of independent structural $q$-modes to describe the thermodynamics and the related underlying dynamics of the fluid -- as stated in the general expression for the free energy given in Eq. (2)  -- breaks down as a more complex wavevector-dependence of the collective transport is involved \cite{chaikinlubensky}. While both the bulk and confined fluids taken under similar thermodynamic conditions show $q$-dependence at small $q$ that does not obey De Gennes narrowing, departure remains small for the latter as compared to the former. Indeed, as shown in Fig. 5(b), the product $D_\textrm{T} \times S(q)$ remains of the order of 1 at all $q$ while it varies by almost a factor of 10 for the bulk fluid in the low $q$ range.

The $q$-dependence observed in $D_\textrm{T}(q)$ at small $q$ (\textit{i.e.} which does not verify De Gennes narrowing) is governed by dynamical contributions that are missing in the thermodynamical approach based on the free energy expression in Eq. (2) combined with simple transport relations in Eqs. (3). In more detail, such terms describe the wavevector dependence of the overall fluid dynamics (\textit{i.e.} dynamical coupling between molecules and their surrounding whose impact depends on $q$). While this $q$-dependence is dominated by de Gennes narrowing in the vicinity of the correlation peak in the structure factor, it is convoluted by a function that evolves on larger -- typically  supramolecular -- length scales. Such additional effects are well known in the case of solvent/solute systems such as suspensions, mixtures, etc., where they are described by the hydrodynamic function $H(q) = D_\textrm{T}(q)S(q)/D_0$ (this function converges to unity at vanishing wavectors $q \to 0$ as $D_\textrm{T} = D_0/S_0$ in the macroscopic limit). Here, we  adopt the same notation with the understanding that $H(q)$ refers to collective dynamical terms that give rise to the additional wavevector-dependence in the fluid dynamics (\textit{i.e.} beyond the structure/dynamics relationship corresponding to De Gennes narrowing). As shown in \textcolor{blue}{\textit{SI Appendix}, Fig. S8}, $H(q)$ shows variations with $q$ for both the bulk and confined fluids in regions outside that corresponding to De Gennes narrowing. For reasons already commented above, both $H(q)$ for the bulk and confined fluids vary with $q$ at large $q$ (\textit{i.e.} $q \gtrsim 2$ \AA$^{-1}$) as the mesoscopic equations that define Onsager's transport coefficients do not hold at such submolecular lengthscales. In the intermediate $q$ range where the concept of De Gennes narrowing applies, $H(q)$ remains constant as expected as the dynamics at the corresponding lengthscales for the bulk and confined fluids are governed by their structure factor $S(q)$. Finally, both $H(q)$ for the bulk and confined fluids vary with $q$ at small $q$, which indicates that their dynamics exhibit additional wavevector-dependence with respect to De Gennes narrowing. The amplitude of these variations with respect to $H(q) = 1$ is found to be much larger for the bulk fluid than the confined fluid, therefore showing that such collective interactions are more important for the former than the latter. This result is attributed to the strong fluid/solid interactions at play in nanoconfinement which screen collective dynamical interactions within the fluid and, hence, decrease the associated $q$-dependence of collective transport. 
The less predominant role of collective effects in the nanoconfined fluid dynamics is consistent with available literature  on fluids in extreme confinement which suggests that collective diffusivity becomes closer to the self-diffusivity (\textit{e.g.} \cite{falk_2015}). In any case, as expected, for both the bulk and confined fluids, at vanishing wavevectors $q \to 0$ (macroscopic scale), the hydrodynamic interactions $H(q)$ become scale-independent as the structure factor $S(q)$ and transport diffusivity $D_\textrm{T}(q)$ become constant. As shown in \textcolor{blue}{\textit{SI Appendix}, Fig. S9} in which a log-log scale representation is used, despite the severe confinement and strong underlying periodic structure imposed by the nanoporous zeolite, the asymptotic macroscopic limit is reached for the confined fluid at similar wavevectors than for the bulk -- typically $q \lesssim 0.05$ \AA$^{-1}$. While considering non-simple fluids (\textit{i.e.} with an underlying molecular structure involving intramolecular modes) can lead to more complex patterns in the structure factor and the $q$-dependent diffusivity, we expect De Gennes narrowing to remain applicable. Similarly, considering that we selected a host confining structure with a non-trivial pore network, the observations reported in the present paper should also apply to materials with complex pore geometries/distributions. In this context, while further study is needed to address the impact of fluid complexity or disordered pore networks on De Gennes narrowing, the extended validity of this simple concept for a nanoconfined fluid is a promising lead to rationalize transport in nanoporous media.\\

    \begin{figure*}
        \centering
        \includegraphics[width=13.5cm]{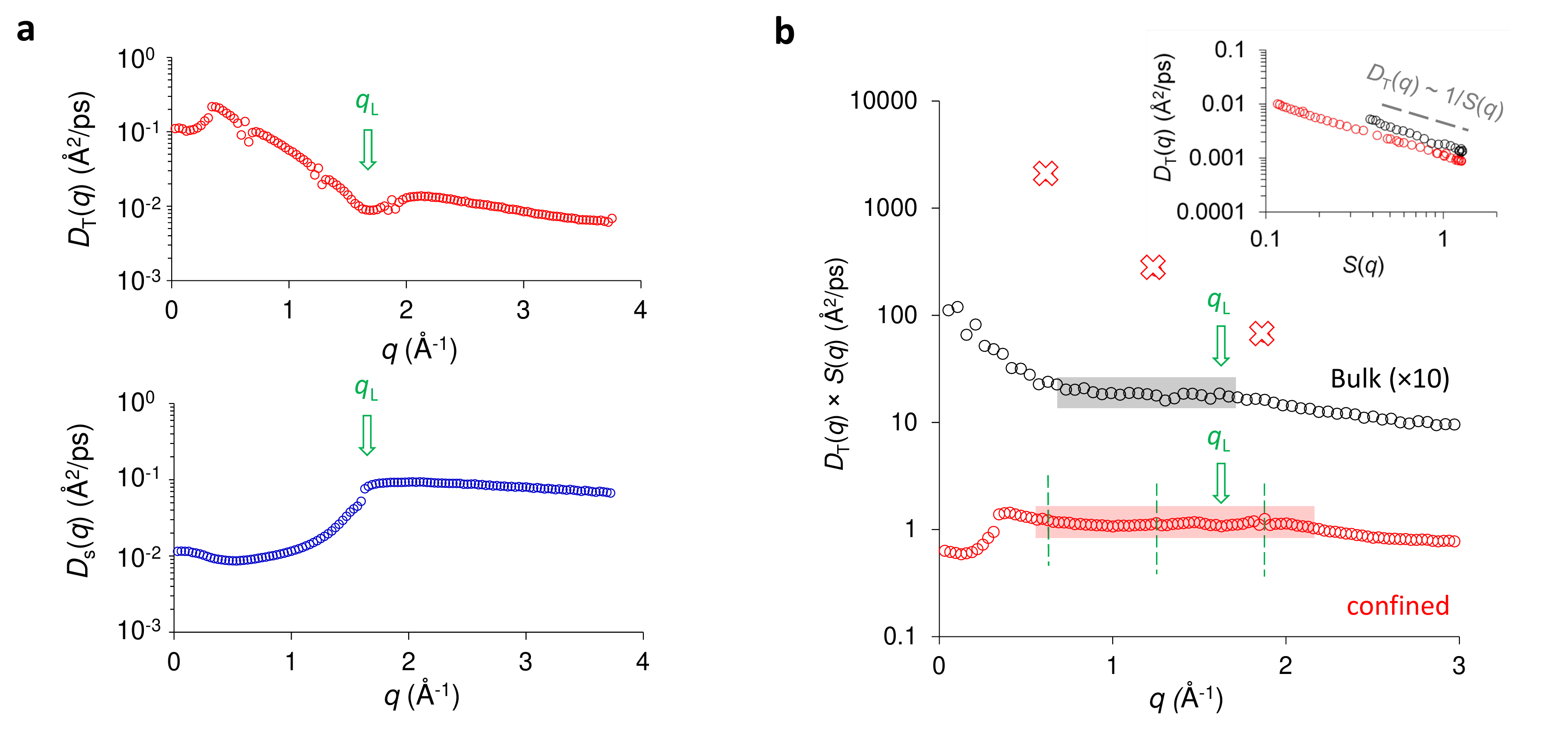}
  \caption{\textbf{De Gennes narrowing in a molecularly confined fluid.} \textbf{a}, Self-diffusivity $D_\textrm{s}(q)$ (bottom) and transport diffusivity $D_\textrm{T}$ (top) as a function of the wavevector $q$ for for methane adsorbed in silicalite-1 zeolite at $T = $ 300 K ($n_\textrm{a} \sim 20$ CH$_4$/uc). The green down arrow indicates the position of the liquid correlation peak $q_\textrm{L}$ in the structure factor $S(q)$. \textbf{b}, Assessment of De Gennes narrowing by plotting $D_\textrm{T}(q) \times S(q)$ as a function of the wavevector $q$. The red circles are for methane confined within silicalite-1 zeolite with a loading $n_\textrm{a} \sim 20$ CH$_4$/uc while the black circles are for bulk methane at the same temperature and a similar density per unit volume.  Note that the data for bulk methane has been multiplied by a factor 10 for the sake of visibility. For each data series, the red and black shaded areas indicate the wavevector range where De Gennes narrowing is observed while the green down arrow indicates the position of the liquid correlation peak $q_\textrm{L}$. As discussed in the text, the big crosses show the effect of taking $S(q)$ instead $\Tilde{S}(q)$ for Bragg conditions (\textit{i.e.} $q = q_\textrm{B}$). The inset shows in a log-log scale $D_\textrm{T}(q)$ as a function of $S(q)$ for wavevectors $q$ corresponding to the red and black shaded areas in the main figure.}

        \label{fig6}
    \end{figure*}

\section*{Discussion}

\noindent The agreement between the simulated data and theoretical predictions indicates that the concept of \textit{De Gennes narrowing} provides a relevant and accurate formalism to describe the collective transport of fluids confined at the molecular scale. Despite severe confinement and strong adsorption effects, which are inherent to fluids in such extreme environments, we find that this simple framework captures the wavevector-dependent dynamics of the confined fluid with a predicting power that surpasses what is observed for the bulk fluid under similar conditions. In this context, the fact that confinement and surface interactions screen the collective dynamical coupling in the nanoconfined fluid renders the theory of De Gennes narrowing even more relevant and applicable. These findings are even more remarkable as the confined fluid displays additional specific features beyond well-known confinement and adsorption effects. First, the confined molecules obey a marked intermittent dynamics as they switch regularly between adsorption at the pore surface and relocation through the pores. Second, the distribution of molecules within the pore space displays strong structural features corresponding to the fingerprint of the host nanoporous material. Overall, while the behavior of nanoconfined fluids often challenges existing framework with novel phenomena being unravelled regularly, our results suggest that the microscopic description at the root of De Gennes narrowing is a cornerstone to build molecular and upscaling approaches of transport in nanoporous media. In fact, beyond the important example of collective dynamics in confined fluids, this work illustrates the power of statistical mechanics approaches to capture the interplay between adsorption and transport as well as the breakdown of continuum level descriptions (\textit{e.g.} hydrodynamics) at the nanoscale. 

From a fundamental viewpoint, the present work addresses a knowledge gap in the realm of nanoconfined fluids by considering a microscopic model for the collective diffusivity and, hence, the permeability. Moreover, from a more practical viewpoint, our findings pave the way for the design of novel applications and processes by considering the rich and complex dynamics of confined fluids at different length and timescales. In this context, we note that the vast majority of experimental and theoretical works on nanoconfined fluids only considers the macroscopic limit. Yet, at this scale, many intriguing and complex phenomena occuring at smaller scales are already averaged out into effective parameters and constants. As a result, it is intrinsically very difficult to unravel the complexity and richness of phenomena involved in nanoconfined fluids from macroscopic experiments (unless specific experiments are developed and carried out such as using nanofluidic devices, surface force apparatuses, etc.). In contrast, unravelling the wavevector-dependent dynamics of nanoconfined fluids is required to understand the possible coupling between fluid transport and monitoring or stimulation techniques at a given length scale. This includes the use of electromagnetic waves such as in microwave-assisted catalysis but also mechanical solicitation to promote transport across nanoporous membranes.

\matmethods{
		\subsection*{Molecular models}
        
        The zeolite, which consists of pure silica MFI crystalline structure, was built from the International Zeolite Association database (IZA)\cite{iza}. The parameters for orthorhombic silicalite-1 are $a=20.09$ \AA{}, $b=19.738$ \AA{} and $c=13.142$ \AA{} [Fig. 1(a)]. The  3D porous network in this zeolite consists of straight channels in the $y$ direction (circular opening $\sim$ 0.55 nm) and sinusoidal, i.e. zigzag, channels in the $xz$ plane (elliptical opening $\sim$ 0.51–0.55 nm). To avoid finite size effects, a silicalite-1 supercell made up of $2 \times 2 \times 2$ cells along the $x$, $y$ and $z$ directions was built by replicating the crystallographic unit cell and periodic boundary conditions were applied. The zeolite flexibility was taken into account using the force field by Vlugt and Schenk in which harmonic potentials are used to describe the Si-O bonds but also fictive bonds between O atoms connected to the same Si atom \cite{vlugt_2002}. As a prototypical simple fluid, methane was considered in this study using a united-atom model in which the methane molecule is described as a single Lennard-Jones sphere \cite{ungerer_2000}. The fluid/zeolite intermolecular interactions are also described using Lennard-Jones potentials (however, no Lennard-Jones potential between CH$_4$ and Si atoms was considered as dispersion interactions with Si can be neglected owing to their small polarizability). The Lennard-Jones parameters are $\sigma=3.73$ \AA{}, $\varepsilon$/k$_b$ = 147.9 K for the CH$_4$-CH$_4$ interaction \cite{ungerer_2000} and $\sigma=3.214$ \AA{}, $\varepsilon$/k$_b$ = 133.2 K for the CH$_4$-O interaction \cite{kar_2001}. All interactions were considered with cut-off $r_c \sim 13$ \AA.

		\subsection*{Grand Canonical Monte Carlo}
		
        All simulations are carried out using LAMMPS simulation package \cite{plimpton_1995} (version OpenMPI 1.8.1). As described in detail in \cite{kellouai_2022}, the fluid adsorption isotherm in silicalite-1 was determined using Grand Canonical Monte Carlo (GCMC) simulations. With this statistical mechanics approach, the zeolite is set in contact with a fictitious bulk reservoir of methane molecules that imposes its temperature $T$ and  chemical potential $\mu$ \cite{frenkelsmit}. Considering the pressure and temperature conditions considered in this work, the chemical potential $\mu$ can be estimated readily from the pressure $P$ using the ideal gas equation of state, $\mu = k_\textrm{B}T \ln [P \Lambda^3/k_\textrm{B}T]$ where $\Lambda$ is the De Broglie thermal wavelength while $k_\textrm{B}$ is Boltzmann constant. Once equilibrium is reached, the adsorbed amount $n_\textrm{a}$ of methane in silicalite-1 is assessed from the average number of molecules  $\langle N \rangle$. To ensure efficient sampling for these simulations performed in the Grand Canonical ensemble, the Metropolis algorithm is used where methane molecules are translated, inserted or removed randomly (Monte Carlo moves) with an acceptance probability defined as the ratio of the Boltzmann factors before and after the attempted move. 
        
        \subsection*{Molecular Dynamics} 
        
        All molecular dynamics simulations were carried out using LAMMPS  \cite{plimpton_1995}. Starting from well-equilibrated configurations obtained using Grand Canonical Monte Carlo simulations, both equilibrium and non-equilibrium molecular dynamics  were performed to determine the self and collective diffusion properties of the confined fluid at different pressures. All MD simulations were performed in the canonical ensemble (\textit{i.e.} constant number of molecules $N$, volume $V$ and temperature $T$) using a Nosé-Hoover thermostat. The simulations were performed for at least 10 ns -- this simulation time was found to be large enough to allow molecules to explore distances larger than 10 nm and, more importantly, to reach the Fickian regime. The time step used to integrate Newton's equations of motion was set to 1 fs and the molecular configurations were stored every 1 ps along the trajectory (more frequent storing was considered when needed to probe short time phenomena). For the non-equilibrium molecular dynamics simulations, a similar strategy with the same time constants was employed but collective transport was investigated by inducing a steady flow rate $\overline{v}$ through a pressure gradient  $\nabla \mu$. In practice,  $\nabla \mu$  was applied by imposing on each molecule a force per unit volume  $f = -\nabla \mu$.  While the force applied was small enough to ensure that the system remains in the linear response regime, \textit{i.e.} $\overline{v} \propto \nabla \mu$, it was checked that the temperature remains equal to that predicted from the equipartition theorem $\overline{v}^2 \propto T$. Moreover, to avoid any unphysical coupling between the Nosé-Hoover thermostat employed to control the temperature and the driving force inducing the flow, the instantaneous temperature when applying the thermostat was corrected to only account for the velocities in the directions perpendicular to the flow. 
        
} 

\showmatmethods{} 

\acknow{W. K. thanks the Fédération Française de Diffusion Neutronique 2FDN (FR2004). Calculations were performed using the GRICAD infrastructure (https://gricad.univ-grenoble-alpes.fr), which is supported by the Rhône-Alpes region (GRANT CPER07-13 CIRA) and the Equip@Meso
project (reference ANR-10-EQPX-29-01) of the programme
Investissements d’Avenir supervised by the French Research
Agency. We wish to thank D. Constantin for useful exchanges on De Gennes Narrowing.}

\showacknow{}

\end{document}